\documentclass[pdflatex,sn-mathphys-num]{sn-jnl}

\usepackage{graphicx}%
\usepackage{multirow}%
\usepackage{amsmath,amssymb,amsfonts}%
\usepackage{amsthm}%
\usepackage{mathrsfs}%
\usepackage[title]{appendix}%
\usepackage{xcolor}%
\usepackage{textcomp}%
\usepackage{manyfoot}%
\usepackage{booktabs}%
\usepackage{algorithm}%
\usepackage{algorithmicx}%
\usepackage{algpseudocode}%
\usepackage{listings}%
\usepackage{ulem}
\usepackage{amsmath,amsfonts}
\usepackage{braket}
\usepackage{geometry}
\usepackage{booktabs}
\usepackage{tabularx}
\usepackage{caption}
\usepackage{tabularx}

\usepackage{array}
\usepackage{adjustbox}
\usepackage{multirow}

\theoremstyle{thmstyleone}%
\theoremstyle{thmstyletwo}%

\theoremstyle{thmstylethree}%

\begin{document}

\title[Article Title]{Towards Universal Material Property Prediction with Deep Learning and Single-Descriptor electronic Density}


\author[1]{\fnm{Feng} \sur{Chen}}

\author[1]{\fnm{Shu} \sur{Li}}

\author[2]{\fnm{Xin} \sur{Chen}}

\author[1]{\fnm{Dennis} \sur{Wong}}

\author*[3]{\fnm{Biplab} \sur{Sanyal}} \email{biplab.sanyal@physics.uu.se}

\author*[1]{\fnm{Duo} \sur{Wang}}\email{duo.wang@mpu.edu.mo} 


\affil[1]{\orgdiv{Faculty of Applied Sciences}, \orgname{Macao Polytechnic University}, \city{Macao SAR}, \postcode{999078}, \country{China}}

\affil[2]{\orgdiv{Thermal Science Research Center}, \orgname{Shandong Institute of Advanced Technology}, \orgaddress{\city{Jinan}, \postcode{250100}, \country{China}}}

\affil[3]{\orgdiv{Department of Physics and Astronomy}, \orgname{Uppsala University}, \city{Uppsala}, \postcode{75120}, \country{Sweden}}




\abstract{

Owing to its high scalability and computational efficiency, machine learning methods have been increasingly integrated into various scientific research domains, including \textit{ab initio}-based materials design. It has been demonstrated that, by incorporating modern machine learning algorithms, one can predict material properties with practically acceptable accuracy. However, one of the most significant limitations that restrict the widespread application of machine learning is its lack of transferability, as a given framework is typically applicable only to a specific property. 
The origin of this limitation is rooted in the fact that a material's properties are determined by multiple degrees of freedom -- and their complex interplay -- associated with nuclei and electrons, such as atomic type, structural symmetry, and the number and quantum states of the valence electrons, among others.
The inherent complexity rules out the possibility of a single machine learning framework providing a full description of these critical quantities.
In this paper, 
we develop a universal machine learning framework based solely on a physically grounded and theoretically rigorous descriptor -- electronic charge density.
Our framework not only enables accurate prediction of eight different material properties (with R$^2$ values up to 0.94), but also demonstrates outstanding multi-task learning capability, as prediction accuracy improves when more target properties are incorporated into a single training process, thereby indicating excellent transferability.
These results represent a significant step toward realizing the long-standing goal of a universal machine learning framework 
for the unified prediction of all material properties.

}






\maketitle



The rapid advancement of machine learning (ML) and deep learning (DL) technologies has opened up unprecedented opportunities in materials science. By leveraging their powerful capabilities in pattern recognition and high-dimensional data processing, these techniques are successfully used to accelerate the discovery and design of novel materials. Unlike traditional approaches reliant on time-consuming experiments or computationally expensive simulations, ML/DL-driven frameworks have demonstrated the ability to efficiently and accurately predict a wide range of materials properties.
However, in spite of the rapid advancement of conventional ML/DL models aimed at either improving the prediction accuracy for a specific property or extending the models to previously unexplored material properties, 
one of the most critical questions in the field remains unanswered: 
can a universal model be constructed to accurately predict all material properties within a unified framework?

In general, evaluating an ML/DL-driven model involves two key aspects: accuracy, which reflects how well the model-predicted results fit known data; and transferability, which characterizes its ability to generalize across different material properties.
Approaches to improving both aspects are inherently rooted in the ML workflow itself, 
which, in practice, maps input datasets to material property outputs via either regression or classification algorithms.
Apart from the conventional approach that focuses on improving the algorithm that technically links 
input and output,
a more physically grounded alternative
lies in refining the selection of a descriptor -- one that, ideally, 
captures the essential features of the material and is closely related to the property of interest.
In this context, model accuracy depends on whether the chosen descriptor can effectively represent the target property,
whereas transferability depends on the extent to which
multiple target properties are intrinsically correlated and 
can be uniformly mapped to the same descriptor.
In principle, 
the more direct the representation and the stronger the correlation embodied by the descriptor, 
the better the resulting ML model performance.

In practice, 
a descriptor encodes characteristic patterns of the material that are identifiable by ML/DL algorithms and assumed to be directly related to the emergence of the target property.
These patterns can be automatically identified by the ML/DL algorithm through an approach known as feature engineering. Although this method has been successfully applied to property prediction with satisfactory accuracy\cite{jiang2021topological,he2021machine} and relatively limited transferability\cite{liu2020leverage}, it still suffers from fundamental limitations that are difficult to overcome.
For example, the descriptors extracted in this way are often merely numerical values without clear physical meaning, making the entire framework resemble a black-box model that offers little insight into the underlying physical mechanisms and thereby limits further interpretability or refinement.
On the other hand, fundamental physical attributes of materials -- such as crystal structure, atomic species, and chemical composition -- are widely considered natural descriptors for material representation, and have thus been widely used in modern ML/DL frameworks.
Recent developments in this regard have primarily focused on constructing potentially more accurate or detailed parameterizations of these quantities from diverse perspectives -- such as bond length and angle\cite{fang2025machine, madani2025accelerating,grunert2024deep}, symmetry consideration (e.g., group theory)\cite{talapatra2023band,zubatyuk2019accurate,broyles2024structure,koker2024higher}, XRD spectra data\cite{kharel2025omnixas}, electronic configuration\cite{talapatra2023band,zubatyuk2019accurate,broyles2024structure,karamad2020orbital}, ionization energy\cite{talapatra2023band}, electronegativity\cite{li2024high}, and Kohn-Sham energy-- by employing a variety of machine learning algorithms. 
In addition, it is natural to expect that incorporating more such physical quantities into the descriptor leads to a more complete representation of the material and, consequently, to more reliable property predictions. While recent studies prove that this approach may improve predictive accuracy, a universal ML framework that achieves acceptable transferability across diverse properties remains elusive.
This can be partially attributed to limitations in the size, diversity, and accuracy of available datasets, 
as well as the inherent complexity involved in mapping input data to output properties -- 
such as uncertainties in experimental data acquisition, theoretical approximations introduced during Hamiltonian construction (e.g., the single-particle approximation), and the resulting numerical errors in solving it, i.e., the Kohn-Sham Hamiltonian. 

In recent years, the importance of electronic charge density has been emphasized and successfully incorporated into ML/DL frameworks. 
According to the Hohenberg-Kohn theorem\cite{hohenberg1964inhomogeneous}, the ground-state wavefunction of a material --i.e., the solution of the Schr\"odinger equation - is in one-to-one correspondence with its real-space electronic charge density. This makes the latter a physically grounded and computationally feasible descriptor for machine learning applications. 
More importantly, since the electronic density is the direct outcome of solving the Kohn-Sham Hamiltonian, it resolves problem by skip the complex mapping process and provides a direct correlation with all the materials' properties.
A milestone
work in 2021 marked the first use of electronic density as the output in a machine learning framework to train a functional, thereby pushing the frontiers of density functional theory (DFT) by enabling a more accurate solution of the Kohn-Sham Hamiltonian\cite{kirkpatrick2021pushing}. More recently, several pioneering studies have explored the use of electronic density as a descriptor; however, these efforts have typically been limited to materials with specific crystal structure (e.g., BCC) or targeted properties (e.g., total energy and bulk modulus)\cite{saha2023electron,arora2022charge,mirzaee2024elastic,ray2025lean}, due to the limited size of manually curated datasets. A comprehensive exploration of the application of electronic charge density in ML/DL framework is essential for advancing modern materials research.
In this study, 
we use electronic density curated from the Materials Project as a unique descriptor for predicting eight different ground-state material properties. 
To fully recognize the rich information contained in the electronic density data, 
we first normalized the 3D data into a series of image snapshots along $z$-direction, and then employed Multi-Scale Attention-Based 3D Convolutional Neural Network (MSA-3DCNN) to extract features suitable for machine learning and thereby establish a mapping to the target properties. 
We adopted both single-task and multi-task learning approaches in our DL model -- with average R2 value of 0.66 and 0.78, respectively -- These results first confirm the feasibility of our proposed approach for predicting material properties based solely on electronic charge density. Furthermore, they demonstrate that multi-task learning significantly enhances prediction accuracy across different properties. Overall, the model exhibits not only high accuracy but also excellent transferability, making it a promising framework for universal property prediction in materials science.


\section*{RESULTS}\label{sec2}

\subsection*{3DCNN-based
workflow for Material Property Prediction from Electronic Charge Density}


\subsubsection*{Model training optimization
}

Depending on the nature of the data and the specific task at hand, different DL models can be constructed.
With respect to the three-dimensional (3D) electronic density data, three models -- PointNet\cite{qi2017pointnet}, Sparse 3D Convolutional Neural Networks (Sparse 3D CNNs)\cite{graham2014spatially,liu2015sparse,wang2017cnn}, and 3D convolutional neural network (3D CNN)\cite{tran2015learning} -- have been successfully applied; however, all of them exhibit critical limitations.
For example, PointNet discretizes the continuous electronic density field into unordered point sets, which effectively reduces computational complexity; however, its heavy reliance on global feature aggregation limits its ability to capture fine-grained local variations in the charge density, such as electron accumulation near chemical bonds.
In addition, it has been shown to exhibit insufficient sensitivity to weak signal regions.
All of these subtle features are essential in determining a material's electronic structure and, thereby, its overall properties. 
Similarly, sparse 3D CNNs improve computational efficiency by ignoring regions with zero-values. These networks were originally designed to handle sparse input data, but not for 3D charge density. Application of this network to studies of charge density inevitably results in downsampling of the data, thereby diminishing the prediction accuracy.
On the other hand, 3D CNN\cite{tran2015learning} is among the most practical and effective DL model for processing data like 3D charge density, as it retain the spatial consistency of the data in real space and inherently capture subtle local feature correlations through it 3D convolution operations.
However, there are two major limitations of this network: the first concerns memory issues during training; the second, and more importantly,  lies in its requirement that all training data must share a unified dimension. This latter condition is not met for charge density generated by first-principles calculations, since their dimensions are directly connected to the lattice parameters in Cartesian coordinates. This makes the data material-dependent and impossible to pre-align without drastically losing computational accuracy.
In this work, we have successfully addressed these problems by converting the 3D matrix data into image representations and by adopting a well-designed interpolation scheme. To the best of knowledge, this is the first time that this combination of procedure has been applied to condensed matter systems. A schematic illustration is provided in Fig. 1(a), and further details will be discussed in the following subsection.

\begin{figure}[H]
    \centering
    \includegraphics[width=0.9\linewidth]{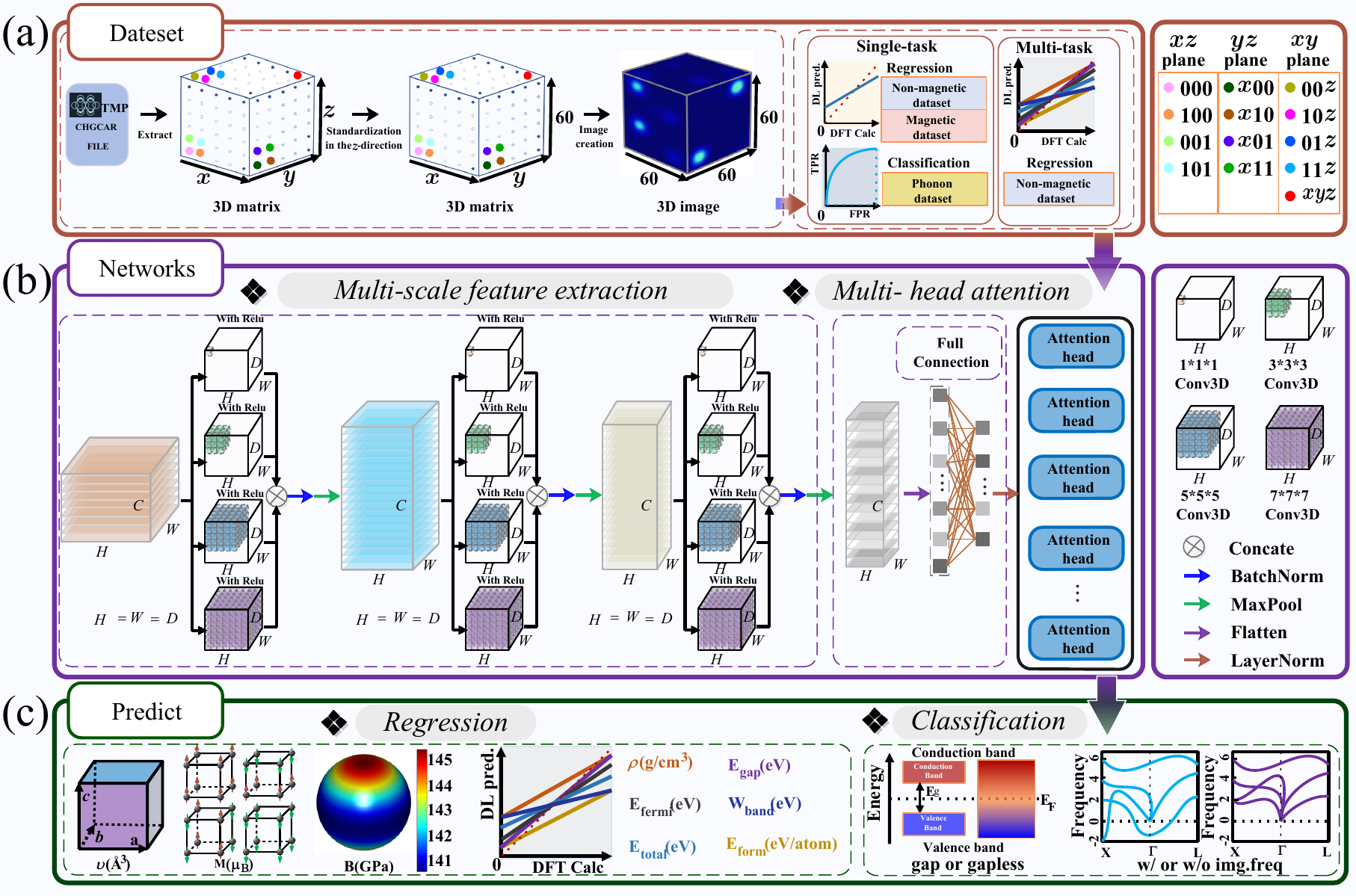}
    \caption{Schematic illustration of the overall workflow of the MSA-3DCNN model and its main components. (a) Data preparation process. (b) The main body of MSA-3DCNN workflow, comprising two major blocks: multi-scale convolution and multi-head attention mechanisms. (c) Properties of interest considered in this study.}
    \label{21}
\end{figure}

\subsubsection*{Charge density data standardization and image representation}

Our dataset is curated from the Materials Project online platform, where all simulations were performed using VASP, and the corresponding charge density data are written in CHGCAR files, representing as three-dimensional $(x,y,z)$ matrices.
For a given material, the data dimensions are determined by the lattice parameters along the three Cartesian directions and the Fast Fourier Transform (FFT) grid settings along the corresponding axes. 
The statistical distribution of dimensional variations across materials in our dataset is shown in Fig. 2(a), where the majority of values along the three directions range from 50 to 200, with a peak around 60.
Since it represents the real-space electronic charge density, a straightforward idea is to visualize it in an image form, as shown in Fig. 2(b). Thereby, we propose a two-step procedure to standardize this 3D image: First, normalize the z-dimension to 60 grid points by linearly interpolating data between neighboring grid points. Second, standardize the in-plane dimensions to $60\times 60$ by extracting in-plane slices for each $z$ value and converting them into 2D images with $x$- and $y$-dimensions fixed at 60.

Taking Ti$_4$O$_8$ as an example, the original data dimensions are $(72,80,84)$, corresponding to grid spacing of 0.0139. 0.0125, and 0.0019 {\AA} along the $x$, $y$, and $z$ directions, respevtively. A 3D visualization and a representative 2D slice at $x=24$ are shown in the two left panels of Fig. 2(c). 
By applying linear interpolation, the number of grid points along the $z$ direction was reduced from 84 to 60, and the spacing between neighboring grid points increased from 0.0019 to 0.0167 {\AA}, while the data along the other two directions remained unchanged. 
The resultant charge density is shown in the right panel, where the variations remain identical before and after interpolation.
This demonstrates that the interpolation process effectively preserves the essential features of the original data while simultaneously
achieving a standardization to the form $(x,y,60)$.
As the second step of standardization, the grids along the $x$- and $y$-directions are unified by extracting 60 slices of data (shown in Fig. 2(a), left panel), each corresponding to a specific $z$ value. These sixty $(x,y,1)$ matrices are then converted into 2D images with a normalized size of $(60,60,1)$ using the Python package matplotlib. The images are in color format, meaning that three separate channels (red, green, and blue) are required when processed by a image analysis algorithm. 
Since the values in the original matrices represent only the magnitude of the charge density, and the colors in the images add no additional information, we further converted the images into grayscale (shown in Fig. 3(a), right panel). This conversion improves the efficiency of both model training and memory usage.



\begin{figure}
    \centering
    \includegraphics[width=0.85\linewidth]{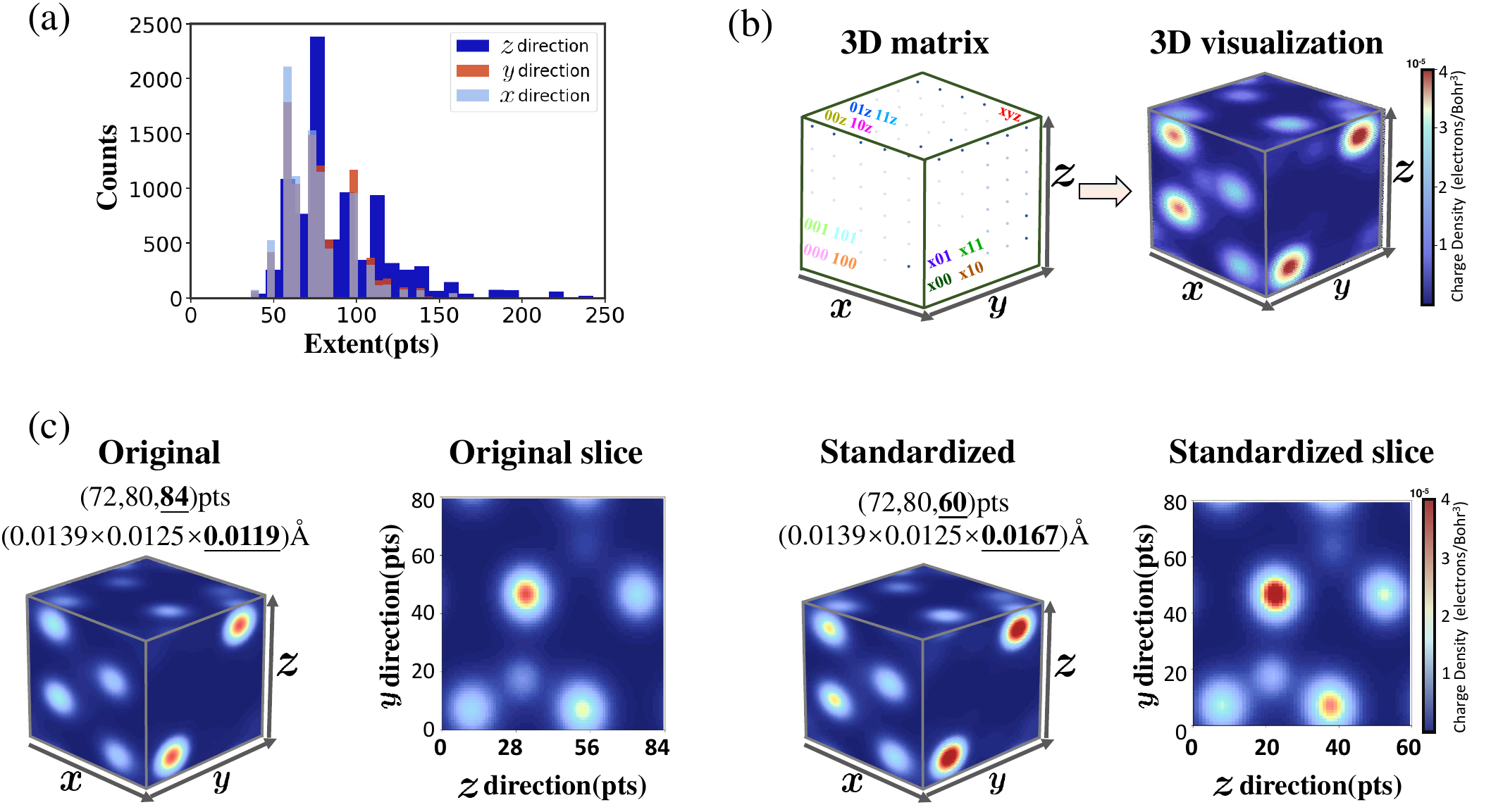}
    \caption{
    \textbf{Charge density data standardization}.
    (a) Illustration of dimensional variations in the dataset, defined as the magnitudes along the $x$-, $y$-, and $z$-direction in the charge density data, respectively.
    (b) Conversion of charge density data from a 3D matrix into a 3D visualization, using Ti$_4$O$_8$ as an example.
    (c) Charge density data before and after standardization along the $z$-direction, using Ti$_4$O$_8$ as an example.
    }
    \label{221}
\end{figure}

\begin{figure}[h]
    \centering
    \includegraphics[width=0.85\linewidth]{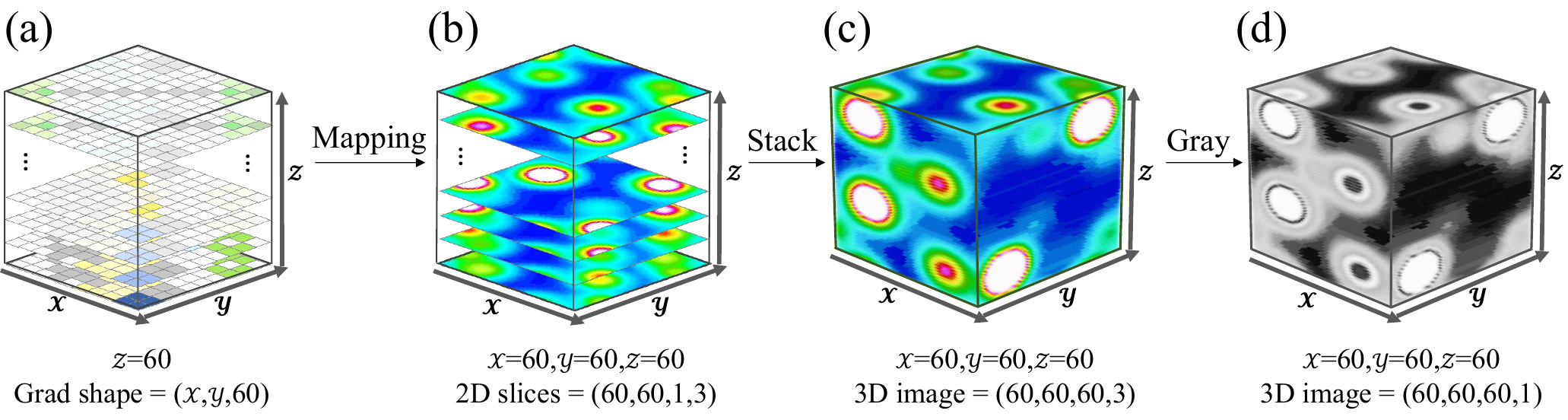}
    \caption{
    \textbf{Image representation of 3D charge density data.}
    (a) 3D charge density grid with $x$- and $y$-scales varying randomly.
    (b) Conversion of 3D charge density data into sixty 2D images, with the x- and y-scales standardized to sixty; each representing a 2D charge density at a specific $z$ value.
    (c) Combined 3D visualization reconstructed from the sixty 2D images. 
    (d) Grayscale version of the previous RGB visualization.
    }
    \label{02}
\end{figure}

\newpage
\subsection*{Model Performance}

In order to validate the transferability and university of our model, we studied a wide variety of material systems, ranging from elemental materials to more complex structures, such as perovskites, XMenes, Heusler alloys, etc. These systems span a wide range of elements, from light to relatively heavy, and cover various structural types such as bulk and low-dimensional form.
For each of these systems, we aimed to cover as many material properties included in our framework as possible.
In this context, we curated data and constructed the following three datasets.
The first is a nonmagnetic dataset, intended for predicting properties such as the bulk modulus B, unit cell volume $v$, density $\rho$, various energy-related quantities (e.g., total energy $\rm E_{total}$, formation energy $\rm E_{form}$, and Fermi energy $\rm E_{fermi}$), as well as electronic structure properties including the band gap and  the bandwidth $\rm W_{band}$.
The second is the phonon dataset, which is used for predicting dynamical stability.
The third is the magnetic dataset, which extends the range of target properties to spin-polarized systems, with a focus on predicting the total magnetic moment per unit cell.

Two conventional deep learning approaches -- classification and regression -- have been employed. The classification model is used to asses material stability and the presence of an energy gap in the electronic band structure, whereas the regression model is responsible for predicting all other quantitative properties.

To further demonstrate the model's transferability, universality, and extensibility rooted in quantum mechanical principles and density functional theory (see Methods), and to enhance predictive accuracy, we conducted both single-task and multi-task learning. The results and their comparison with those reported in recent ML/DL studies are presented in the following subsections.

\subsubsection*{Single-task learning}
For single-task training, we first employed a regression approach to predict nine material properties.
These include three structural-related quantities ($v$, B, $\rho$), five energy-related properties (E$_{\rm total}$, E$_{\rm form}$, E$_{\rm Fermi}$), two electronic structure related properties (E$_{\rm gap}$), W$_{\rm band}$), all based on the NM dataset,and one magnetic property (M), based on the M dataset.
Two key performance metrics, $R^2$ and RMSE, were evaluated to assess the accuracy of the predictions (see Fig. X(a)).
Figure x(b) and Table S1 summarize the $R^2$ values on both the validation and test sets for all eight target properties. The high consistency between them demonstrates the model's stable training and strong generalization performance. Hence, we focus only on the test set results in the following.
Four energy-related properties -- E$_{\rm total}$, E$_{\rm Fermi}$, E$_{\rm form}$, and E$_{\rm gap}$ -- fall in to the second group in terms of predictive performance, with $R^2$ values ranging from 0.52 to 0.59. Among all properties, W$_{\rm band}$ shows the lowest accuracy, with an $R^2$ value of 0.41.
These reductions in accuracy reflect the complex correlations between the corresponding material properties and the real-space charge density. Under the single-task setup, the model is unable to fully capture the relevant electronic features -- in particular, the subtle intensity variations embedded within the continuous, cloud-like distribution of the 3D charge density, which are essential for determining the exact property of interest.
One effective way to address this issue is to incorporate additional input data that contains more complete charge density information.
To this end, we implemented spin-polarized charge density (quadrupled data volume, details in Methods) as the input and performed magnetic moment prediction under the same single-task setup.
As shown in Fig. 4(b), the $R^2$ value in this case reaches 0.75, demonstrating not only the potential for improving accuracy by enriching the input data, but also the feasibility of extending the model to more nontrivial material properties.
Another natural and straightforward approach is to adjust the model architecture to facilitate the extraction of specific electron patterns that can be mapped to the corresponding properties. In this regard, closely related properties may share common features in the charge density, and multi-task training provides a critical framework for addressing such situation -- a topic that will be discussed in  detail in the next subsection.

In addition, we further evaluate our framework by applying a binary classification approach to two properties:
(i) predicting whether a material is metallic or insulating, and (ii) determining whether the material is thermodynamically stable.
Two metrics for evaluating prediction accuracy --  receiver operating characteristic (ROC) and confusion matrix -- are adopted, and the results are shown in Fig. 4(b).
For the first property -- metallic versus insulating -- the area under the ROC curve (AUC) is 0.89 and 0.86 for the validation and test sets, respectively. With an optimal threshold of 0.13, the model achieves an accuracy of 0.83 and 0.86 in correctly predicting the two states, demonstrating high predictive capability of our model.

\begin{figure}[h]
    \centering
    \includegraphics[width=0.9\linewidth]{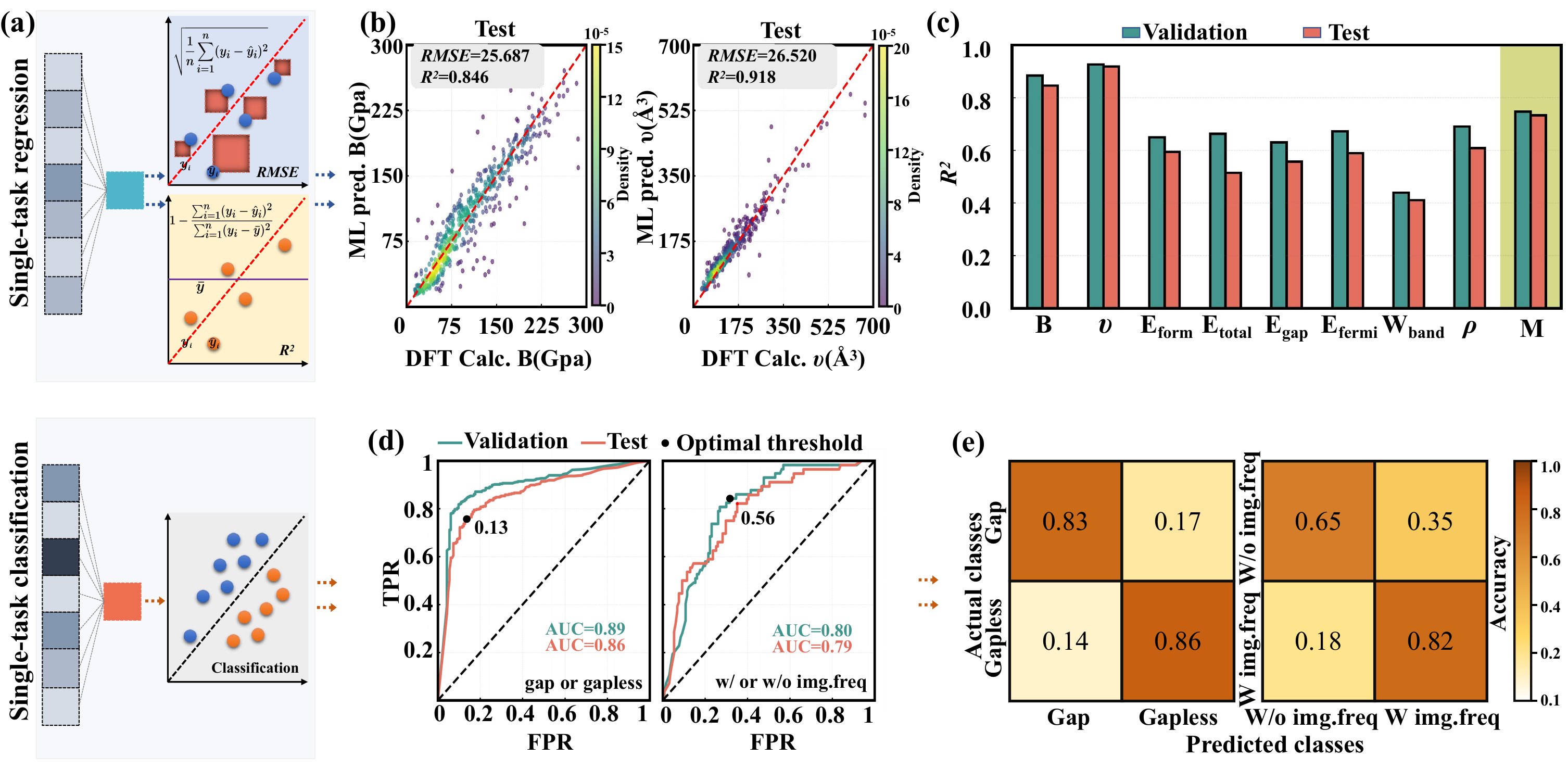}
    \caption{
    \textbf{Results obtained from single-task DL prediction.}
    (a) Results of the regression tasks.
    \textit{Left panel}: Schematic illustration and key evaluation metrics of the regression task. The red dashed diagonal lines represent the true value obtained from DFT calculations. 
    \textit{Middle panel}: Scatter plots of two representative results obtained from our DL model -- bulk modulus and volume -- for the test set. The scale bar on the right shows the data density of the predicted values. 
    \textit{Right panel}: overall prediction performance for nine different target properties; white and light-green backgrounds indicate the results are obtained from non-magnetic and magnetic datasets, respectively.
    (b) Results of the classification tasks. \textit{Left panel}: Schematic illustration of the classification task.
    \textit{Middle panel}: Predicted ROC curves for electronic band gap and dynamic stability classification on both the validation and test sets; the x- and y-axes reprent the false-positive and true-positive rates, respectively. The optimal threshold $\epsilon_{\rm th}$ and the corresponding area under the ROC curve (AUC) are annotated in the plots.
    \textit{Right panel}: Confusion matrices obtained based on the optimal thresholds.
    }
    \label{23}
\end{figure}

\subsubsection*{Multi-task learning
}


It is well known that many properties within a given material are inherently correlated, meaning that one property can be tuned or manipulated through another. 
This intrinsic interdependence forms the physical foundation upon which multi-functional materials are designed.
According to the quantum mechanical principles and the Hohenberg-Kohn theorem (see Methods for details), all material properties are fundamentally determined by the charge density. In other words, these properties are inherently interconnected, with each being mappable to specific patterns within the charge density data. In principle, the stronger the correlation between properties, the more likely they are to be associated with the same underlying patterns.
In this context, we introduce a multi-task learning approach into our framework to jointly learn such interdependent features, with the aim of further improving prediction accuracy and evaluating its transferability and universality.
To systematically examine inter-property dependencies, three property group strategies are employed: (i) Group 1 is formed by randomly clustering the properties; (ii)
Group 2 is based on the Pearson correlation coefficient, which captures linear correlations among the properties of interest; (ii) Group 3 includes all nine properties grouped together, to assess their overall correlation.

Existing studies have demonstrated that physical properties are often interconnected rather than existing in isolation\cite{yankowitz2018dynamic,al2021hydrogen,xing2023structural,khan2022strain}. Here, we are particularly interested in investigating whether capturing the interrelationships between material properties from a data perspective can enhance model performance. Building upon our base model framework, we have modified the output architecture to accommodate multi-task predictions using various task groupings. However, determining the optimal grouping of tasks remains a critical challenge that requires careful consideration. Prevailing wisdom suggests tasks which are similar or share a similar underlying structure may benefit from training together in a multi-task system\cite{caruana1997multitask,fifty2021efficiently}. To systematically examine the impact of task groupings on model performance, we conducted three experiments in Fig. 5(a). First, we implemented a random grouping of tasks; second, we established a grouping through target similarity; and third, we considered the grouping of the entire tasks.


\begin{figure}[h]
    \centering
    \includegraphics[width=0.9\linewidth]{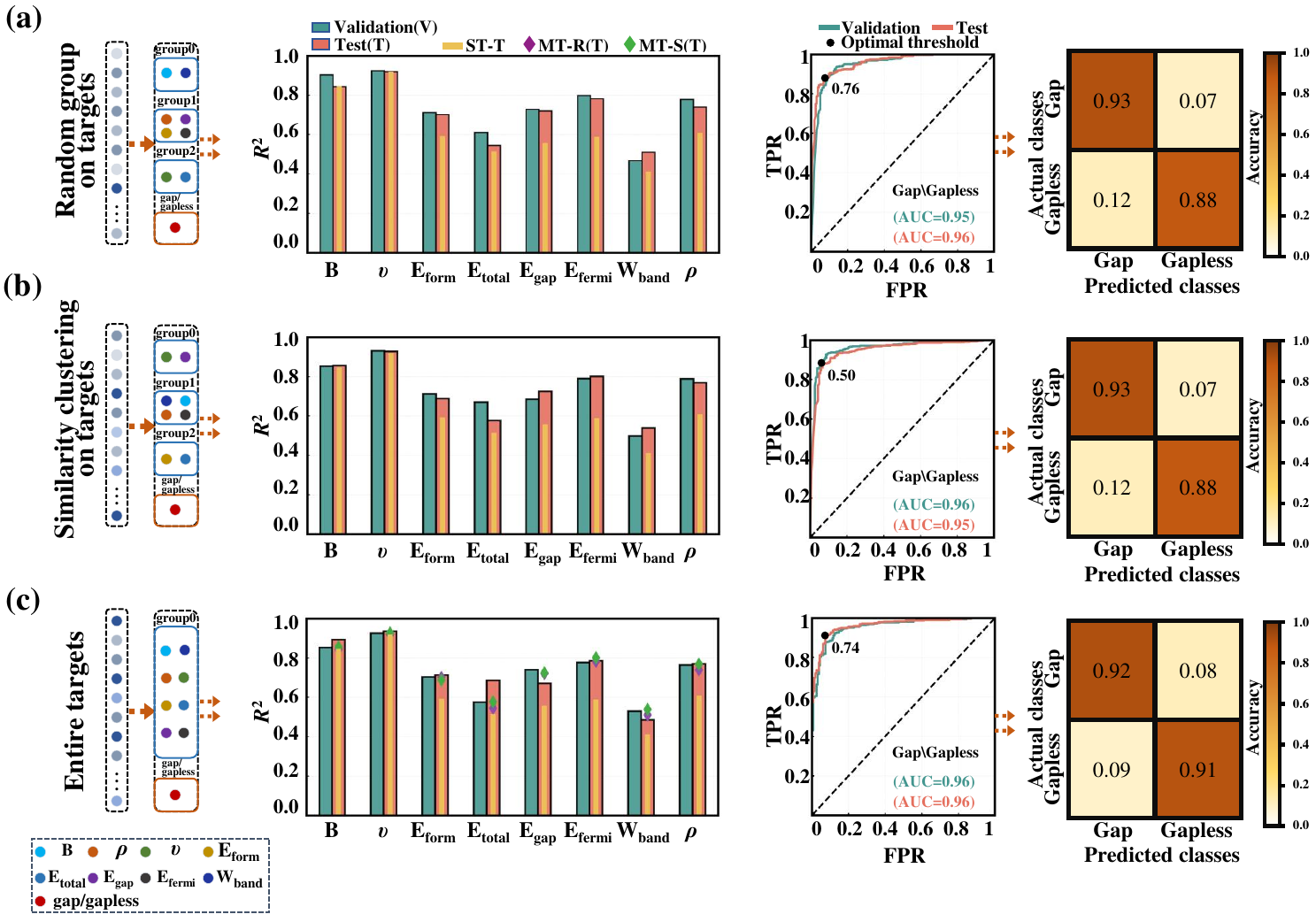}
    \caption{Evaluation of regression and classification performance across different task grouping strategies in a multi-task learning framework. (a) Different task groupings strategies: random grouping (Random group), similarity-based clustering (Similarity clustering) grouping, and training all tasks together (Entire targets). (b) Bar charts depicting the $ R^2 $ scores and performance gaps for different tasks (such as $\text{B}$, $\upsilon$, $\text{E}_\text{form}$, $\text{E}_\text{total}$, $\text{E}_\text{gap}$, $\text{E}_\text{fermi}$, $\text{W}_\text{band}$, and $\rho$ across single-task, random grouping, similarity-based clustering grouping, and entire-targets grouping training strategies. (c) ROC curves illustrating the classification performance for gap and gapless categories, highlighting True Positive Rate (TPR) against False Positive Rate (FPR) at various thresholds, with AUC values indicated for both validation and test sets. (d) Confusion matrices under the best optimal threshold, showcasing accuracy based on task groupings for multi-task learning, with detailed breakdowns for actual versus predicted classes, including gap and gapless categories.}
    \label{05}
\end{figure}

As shown in Fig. 5(b) and Fig. 5(c), the regression and classification prediction results under different task grouping strategies are presented based on the NM dataset. In our random task grouping experiments without relying on prior domain knowledge or explicit attribute relationships, we categorized tasks into regression and classification groups, with further subgroupings in regression tasks ($\text{B}$ and $\text{W}_\text{band}$; $\upsilon$ and $\text{E}_\text{total}$; $\text{E}_\text{form}$, $\rho$, $\text{E}_\text{gap}$, and $\text{E}_\text{fermi}$). Compared with single-task approaches, nearly all properties exhibited higher $R^2$ values except for $\text{B}$, where the performance decline likely stems from negative competitive effects induced by this task grouping strategy. The $R^2$ improvement for $\upsilon$ suggests its prediction has approached the theoretical limit under current data and model configurations. In contrast, properties ($\text{E}_\text{form}$, $\text{E}_\text{total}$, $\text{E}_\text{gap}$, $\text{E}_\text{fermi}$, $\text{W}_\text{band}$, $\rho$) demonstrated significant $R^2$ gains, confirming effective utilization of inter-task correlations. Furthermore, feature sharing enhanced classification performance, achieving 0.96 AUC and 93\% gap identification accuracy (12.05\% improvement over single-task). For the regression, the average $R^2$ reached 0.72, representing a 14\% improvement compared to the result obtained from single-task training.

Next, we grouped tasks based on similarity principles with the objective of further improving model performance. This method employs Pearson correlation-based clustering to partition target properties into groups with statistically correlated behaviors. As shown in the middle section of Fig. 5(a), the clustering results grouped $\text{E}_\text{gap}$ and $\upsilon$ into one cluster, $\text{E}_\text{form}$ and $\text{E}_\text{total}$ into the second cluster, $\text{B}$, $\rho$ ,$\text{W}_\text{band}$ and $\text{E}_\text{fermi}$ into the third cluster. The gap/gapless classification was treated as a separate classification task group. The model demonstrated superior performance when using similarity-based task grouping compared to single-task learning, with improvements observed across all properties - including $\text{B}$ which previously suffered from negative competitive effects under randomly grouped training. For regression tasks, the average improvement $R^2$ of 17.76\% relative to the baselines of the single task increased further by approximately 0.02 compared to the randomly grouped training, representing highly encouraging results. The result supports the conclusion that tasks with lower performances benefit more from multi-task learning with similar targets such as $\text{W}_\text{band}$, $\text{E}_\text{fermi}$, $\text{E}_\text{gap}$, which improve over 30\% compared to the original single-task result. In the gap/gapless classification task, the model achieved an AUC values of 0.95 with a gap identification accuracy of 93\% at $\epsilon_{\text{th,gap}}$ = 0.5. These results further substantiate the inherent
connections between different material properties, demonstrate the effectiveness of similarity-driven task grouping in optimizing feature sharing and capturing complex inter-property relationships among materials, thereby generating meaningful task grouping that enhance predictive performance.


We think that these task grouping methods may still be insufficient for fully optimizing model performance, since our predictions are fundamentally based on electronic density - which reflects electron distribution - and all properties share the same underlying physical principles and interrelationships. Therefore, we grouped all tasks by type, all regression targets were grouped together, while binary gap/gapless classification task was processed separately, as illustrated in the lower panel of Fig. 5(a). The prediction results are presented in the lower panels of Fig. 5(b) and Fig. 5(c). The model demonstrates superior performance across all task groups compared to single task training, with improvements observed across all properties. Regression tasks achieved an average performance improvement of 18\% relative to baselines of a single task, reaching 0.74. However, the improvement becomes marginal compared to the results of similarity-based grouping training, indicating that the strategy of optimizing shared features through task grouping is approaching its prediction limit based on the current number of prediction targets, input data, and the model. This outcome effectively balances the intrinsic correlations among targets with inter-task competition effects. For the gap/gapless classification task, the model achieved an AUC of 0.96 with a gap identification accuracy of 92\% at $\epsilon_{\text{th,gap}} = 0.74$. We have found that these three task grouping strategies exhibit a high degree of reliability in the gap/gapless classification task, with consistent and stable predictive outcomes. This enhanced performance indicates that these strategies effectively capitalize on the relationships between target properties, thereby yielding more accurate and reliable prediction results.

\subsubsection*{Performance Benchmarking with Previous Studies
}

To verify the reliability of the predictive results of our proposed model, we conducted a comprehensive comparative analysis. Our focus was on assessing the predictive accuracy and input data divergence of our model with those state-of-the-art DL models, rather than solely pursuing the extreme accuracy of each property. In selecting the models for comparison, priority was given to advanced models that share the same predictive targets as in this study. Additionally, we aimed to cover representative models of different input feature representation methods to ensure the comprehensiveness and reliability of the comparison. The comparative models included optimal models based on compositional features (such as VAE-based k-NN\cite{ihalage2021analogical}, XGBOOST\cite{elbaz2024density}, Random subspace REPTree\cite{ward2016general}), top-performing convolutional structural feature graph neural network models (such as CGCNN\cite{levamaki2022predicting}, CGCNN-RF\cite{shapera2024machine}, ShotgunCSP\cite{chang2024shotgun}), advanced models incorporating various features after meticulous feature engineering (such as OGCNN\cite{karamad2020orbital}, Hierarchical Down selection\cite{talapatra2023band}), and models using the same features (3D real space electronic density) as input (SDCNN\cite{kim2024deep}).

\begin{table}[h]
\centering
\caption{Comparison of our work with other works.}
\label{tab:comparison}
\renewcommand{\arraystretch}{1.5}
\setlength{\tabcolsep}{0.0pt} 

\tiny 
\begin{tabularx}{\textwidth}{>{\centering\arraybackslash}p{1.1cm}
    >{\centering\arraybackslash}p{1.5cm}
    >{\centering\arraybackslash}p{1.5cm}
    >{\centering\arraybackslash}p{1.3cm}
    >{\centering\arraybackslash}p{1.5cm}
    >{\centering\arraybackslash}p{1.3cm}
    >{\centering\arraybackslash}p{2cm}
    >{\centering\arraybackslash}p{2cm}
    >{\centering\arraybackslash}p{1.8cm}
    >{\centering\arraybackslash}p{1.3cm}
}

\toprule
 & gap/gapless & B(GPa) & $v$(\AA$^3$) & ${\text{E}}_{\text{fermi}}$(eV) & ${\text{E}}_{\text{total}}$(eV) & ${\text{E}}_{\text{gap}}$(eV) & ${\text{E}}_{\text{form}}$(eV/atom) & $\rho$(g/cm$^3$) & ${\text{M}}$($\mu_B$) \\
\midrule

\multirow{4}{*}{\bfseries Other work} 
    & Accuracy-82\%\cite{elbaz2024density} 
    & R$^2$-0.8\cite{mirzaee2024elastic} 
    & R$^2$-0.99\cite{zubatyuk2019accurate} 
    & MAE-0.38\cite{karamad2020orbital} 
    & R$^2$-0.99\cite{fang2025machine} 
    & R$^2$-0.70-0.85\cite{ihalage2021analogical,elbaz2024density,jin2025transformer,talapatra2023band} 
    & R$^2$-0.84-0.98\cite{jin2025transformer,talapatra2023band} 
    & R$^2$-0.68-0.91\cite{nguyen2021predicting} 
    & R$^2$-0.98\cite{fang2025machine} \\
 & Accuracy-95\%\cite{talapatra2023band} 
    & MAE-11.6\cite{levamaki2022predicting} 
    & MAE-49\cite{shapera2024machine} 
    & R$^2$-0.60\cite{massa2024transfer} 
    & -- 
    & MAE-0.28--0.39\cite{kim2024deep,bystrom2024training,ihalage2021analogical,karamad2020orbital} 
    & MAE-0.03-0.488\cite{kim2024deep,ward2016general,karamad2020orbital,chang2024shotgun} 
    & -- 
    & -- \\

\midrule
\multirow{2}{*}{\bfseries Our work} 
    & Accuracy-93\% 
    & R$^2$-0.89 
    & R$^2$-0.94 
    & R$^2$-0.79 
    & R$^2$-0.69 
    & R$^2$-0.73 
    & R$^2$-0.71 
    & R$^2$-0.77 
    & R$^2$-0.75 \\
 &  
    & MAE-14.27 
    & MAE-13.77 
    & MAE-0.77  
    & -- 
    & MAE-0.33 
    & MAE-0.31 
    & --  
    & -- \\


\bottomrule
\multirow{1}{*}{\bfseries Inputdata} 
    &\multicolumn{9}{c}{3D electronic density\cite{kim2024deep,bystrom2024training,mirzaee2024elastic,massa2024transfer}, element or crystal structure features\cite{ihalage2021analogical,levamaki2022predicting,shapera2024machine,elbaz2024density,chang2024shotgun}, multiple sources of features\cite{jin2025transformer,karamad2020orbital,fang2025machine,talapatra2023band,zubatyuk2019accurate}} \\
\end{tabularx}
\end{table}

For the classification task of identifying gaps, our model achieved an accuracy rate of 93\%, outperforming the XGBOOST model\cite{elbaz2024density} achieved an accuracy rate of 82\% and matching the performance of the Hierarchical Down-selection workflow model\cite{talapatra2023band} achieved an accuracy rate of 95\%. Regarding the prediction of regression properties, when the input features were limited to 3D electronic density alone, our model not only significantly outperformed the CLIP neural network\cite{massa2024transfer} which $R^2$ of $\text{E}_\text{fermi}$ is 0.60 and the baseline CNN model\cite{mirzaee2024elastic} which $R^2$ of $\text{B}_\text{GPa}$ is 0.8, but also achieved predictive accuracy comparable to that of the state-of-the-art BP model\cite{bystrom2024training}(MAE of 0.28 eV for $\text{E}_\text{gap}$) and SDCNN model\cite{kim2024deep}(MAE of 0.35 eV for $\text{E}_\text{gap}$ and 0.19 eV for $\text{E}_\text{form}$). It is particularly noteworthy that, when compared with models using element features or crystal structure features as inputs (such as VAE-based k-NN\cite{ihalage2021analogical}, CGCNN\cite{levamaki2022predicting}, CGCNN-RF\cite{shapera2024machine}, XGBOOST\cite{elbaz2024density}, ShotgunCSP\cite{chang2024shotgun}), our model still maintained comparable predictive performance, the result that fully validates the reliability and robustness of our model's predictive outcomes. Regrettably, compared to composite models that integrate multiple sources of features, including crystal structure, elemental features, and electronic structure (such as OGCNN\cite{karamad2020orbital}, AIMNet\cite{zubatyuk2019accurate}, CT-CGCNN\cite{jin2025transformer}, Linear regression\cite{fang2025machine}, Magnetic Monte Carlo simulations\cite{fang2025machine}, etc.), our model slightly lags in regression predictive performance. indicates This that while our model performs well on a single set of features, in future research, we could also consider integrating various features to further enhance the model's predictive capabilities.

\section*{DISCUSSION}



We adopted both single-task and multi-task learning approaches in our DL framework. In the single-task setting, the model achieved the best $R^2$ value of 0.92 and average of 0.63, 
exemplifying the Hohenberg-Kohn theorem by 
highlighting
the critical correlation 
between electronic charge density and the associated
material properties. Furthermore, the prediction accuracy was 
significantly improved
when multiple target properties were combined into a single training process. 
Of particular importance, this improvement 
depends more on the number of
target properties included in the training process, rather than on 
their classification imposed either manually or algorithmically.
More specifically, 
in Task 1 (random grouping of target properties), the average $R^2$ value reached $0.72$, a 14\% improvement over single-task training. 
With the aid of Pearson correlation coefficient-based grouping (Task2), the average $R^2$ further increased by approximately 0.02, reaching 0.74.
Crucially,
when all eight curated properties were grouped and trained together (Task3), the average $R^2$ value rose to 0.74 -- corresponding to a total improvement of 18\% compared to the original single-task result.
These results clearly substantiate the inherent connections between different material properties and strongly support the validity of the HK theorem, which establishes that 
all material properties are functionals of the ground-state electronic density.
Thereby in a deep learning framework, the more of these internal connections and the corresponding characteristic patterns detected by 
the multi-task approach
from the input electronic density, the more accurate the overall material representation, and consequently, the better the prediction accuracy.

It is worth noting that, under the single-task setting, the $R^2$ values for two properties -- volume and bulk modulus -- have already reached 0.92 and 0.84, respectively. With the adoption of the multi-task setting, which allows for shared representations during training, other essential electronic properties, such as Fermi energy, band gap, formation energy, have also achieved $R^2$ values approaching or exceeding 0.70.
This level of accuracy is consistent with the modern machine learning models, yet it is achieved here using a 
more fundamental quantum mechanical quantity -- the electronic electronic density --
as the single descriptor.
More crucially, 
incorporating additional target properties into the model is expected to further enhance prediction accuracy. This gives our framework a distinct advantage in terms of scalability, enabling, in principle, accurate prediction of arbitrary material ground state properties within a unified framework.
With a systematically curated and diverse dataset, the scope of target material property prediction
can be extended to 
more complex quantities, including the magnetic exchange parameters, topological magnetic states, topological spectra (electronic, phonon, magnon), superconductivity, and experimental spectral data (XRD, XPS, \textit{etc}.). Furthermore, it is expected that even more fundamental theoretical quantities -- such as the mean-field Hamiltonian, Kohn-Sham eigenstates, and one-particle density matrix -- will be attainable.
Achieving accurate prediction of some of these quantities is a primary goal of our forthcoming work.
In summary, we have successfully established a universal deep learning framework capable of predicting all ground state material properties with a fundamental quantum mechanical quantity as the unique descriptor, 
constituting a significant step toward the long-standing goal of unified materials property prediction in modern computational material science.

Of equal importance is that our methodology relies heavily on the quality of the data, 
which in turn depends on how the data is generated. 
This is a process that can be systematically improved.
For example, in this study, all electronic density data were curated from the online platform Materials Project, where the calculations were performed in a high-throughput manner using VASP -- a code that employs pseudopotentials and a plane-wave basis set.
Particularly, for a given material, the properties are usually obtained from different simulations, and the same applies to the charge density.
Using data from more systematic and rigorous calculations is expected to enhance prediction accuracy by ensuring higher consistency and reliability in the input features.
On the other hand, using computational codes that employ more sophisticated basis sets -- such as linearized augmented plane wave (LAPW) and linear muffin-tin orbitals (LMTO) -- may further improve prediction accuracy for complex systems, especially for transition metals and their compounds.
Furthermore, in addition to the uniquely employed electronic density input, traditional descriptors commonly used in modern ML/DL frameworks, for example, crystal structure, atomic species, and related features, can be incorporated to further extend the accuracy frontiers demonstrated in this study.

\section*{METHODS}

\subsection*{Theoretical Background}

All materials properties originate from the principles of quantum mechanics, which can be described by the Schr\"odinger equation. The simplest form of a multi-particle system -- non-relativistic and time-independent -- consists of interacting atomic nuclei and their surrounding electrons can be described by the following form 
\begin{equation}
   \hat{H}\ket{\Psi} = E \ket{\Psi},
   \label{Schrodinger equation} 
\end{equation} where $\hat{H}$, $\ket{\Psi}$  and $E$ are the Hamiltonian, wavefunction and energy of
the many-body system, 
respectively. 
For a wavefunction (or the so-called eigenstate) $\ket{\Psi}$, any observable is expressed as an expectation value of an operator $\langle \hat{O} \rangle$
, with the following form 
\begin{equation}
    \langle \hat{O} \rangle=\frac{\bra{\Psi}  \hat{O} \ket{\Psi}}{\braket{\Psi | \Psi}}.
    \label{expectation value}
\end{equation}

The central goal of electronic structure theory is to solve equation~\ref{Schrodinger equation} with sufficient accuracy so that one can predict the diverse properties exhibited by materials according to equation~\ref{expectation value}.
The first and most straightforward approach is to explicitly construct the many-body wavefunction and compute the observables of interest using equation~\ref{expectation value}. However, this approach is severely hindered (circumvented) by the exponential growth of the wavefunction size and, more critically, by the increasing complexity of many-body interactions with respect to system size.
Therefore, obtaining an exact solution to equation~\ref{Schrodinger equation} is practically impossible in the study of real materials.
Instead, an alternative approach is to reformulate the complex many-body problem into an effective single-particle theory based on the electronic charge density, which is amenable to numerical calculations and provides deeper physical insight.
The latter is the so-called Density Functional Theory (DFT), which is based on the Hohenberg-Kohn theorem~[1964] and the Kohn-Sham scheme~[1965]. The former formally (rigorously) established that the ground-state electron density uniquely determines all properties of the system, while the latter provides a practical approach to construct the energy functional based on a given electron density.

The Hohenberg-Kohn theorem states that total energy $E$ of a system with interacting electrons in an external potential -- specifically, the Coulomb potential due to the nuclei in a solid -- is exactly determined as a functional of the ground state electronic density, 
\begin{equation}
\begin{aligned}
    E=E_{\rm HK}[n({\bf r})] & = T[n({\bf r})] + E_{\rm int}[n({\bf r})] + \int n({\bf r})\, v_{\rm ext}({\bf r})\, d^3{\bf r} +E_{II} \\
    & = F_{\rm HK}[n({\bf r})] + \int n({\bf r})\, v_{\rm ext}({\bf r})\, d^3{\bf r} + E_{II},
\end{aligned}
\label{HK functional}
\end{equation}
in which $T[n]$ is the kinetic energy, $E_{\rm int}$ is the electron-electron interaction energy, $v_{\rm ext}$ is the external potential that acts on electrons, and $E_{II}$ is the interaction energy of the nuclei. The newly defined functional $F_{\rm HK}[n]$ includes all internal energies of the interacting electron system.
With electron-density operator defined as
\begin{equation}
    \hat{n}({\bf r}) = \sum_{i=1}^N\, \delta({\bf r}-{\bf r}_i),
\end{equation}
the critical variable, electron density can be obtained as
\begin{equation}
    n({\bf r})=\bra{\Psi} \hat{n}({\bf r})\ket{\Psi}.
\end{equation}

The validation of this theorem can proceed by \textit{reducio ad absurdum}.
Suppose there exist two distinct external potentials, $v_{\rm ext}^{(1)}$ and $v_{\rm ext}^{(2)}$, each associated with a Hamiltonian, $\hat{H}^{(1)}$ and $\hat{H}^{(2)}$, and a corresponding many-body ground-state wavefunction, $\Psi^{(1)}$ and $\Psi^{(2)}$, that yield the same ground-state electron density $n({\bf r})$. Since $\Psi^{(1)}$ is not the ground state of $\hat{H}^{(1)}$, there is
\begin{equation}
E^{(1)}=\bra{\Psi^{(1)}}\hat{H}^{(1)} \ket{\Psi^{(1)}} < \bra{\Psi^{(2)}}\hat{H}^{(1)} \ket{\Psi^{(2)}}.
\end{equation}
The last part/expression/term is the expectation value of $\hat{H}^{(1)}$ with respect to $\Psi^{(2)}$, which can be further written as
\begin{equation}
    \begin{aligned}
        \bra{\Psi^{(2)}} \hat{H}^{(1)} \ket{\Psi^{(2)}} & =  \bra{\Psi^{(2)}} \hat{H}^{(2)} \ket{\Psi^{(2)}} + \bra{\Psi^{(2)}} \hat{H}^{(1)} - \hat{H}^{(2)} \ket{\Psi^{(2)}} \\
        &= E^{(2)} + \int d^3 r \left[v_{\rm ext}^{(1)} - v_{\rm ext}^{(2)} \right] n({\bf r}),
    \end{aligned}
\end{equation}
due to the fact that the Hohenberg Kohn functional $F_{\rm HK}[n]$ must be universal by construction since the two terms included are functionals only of the charge density. So there is 
\begin{equation}
    E^{(1)} < E^{(2)} + \int d^3 r \left[v_{\rm ext}^{(1)} - v_{\rm ext}^{(2)} \right] n({\bf r}).
    \label{conclu1}
\end{equation}

Similarly, applying the same reasoning to $E^{(2)}$ yields 
\begin{equation}
    E^{(2)} < E^{(1)} + \int d^3 r \left[v_{\rm ext}^{(2)} - v_{\rm ext}^{(1)} \right] n({\bf r}),
    \label{conclu2}
\end{equation}
the sum of equations~\ref{conclu1} and~\ref{conclu2} results in an inconsistency 
\begin{equation}
    E^{(1)} + E^{(2)} < E^{(2)} + E^{(1)}.
\end{equation}
Therefore, any given external potential corresponds 
to a unique ground-state density $n({\bf r})$.

In addition, the Hohenberg Kohn theorem further showed that the true ground-state density is the density that minimizes $E[n]$. This variation principle is further clarified by the two-step minimization procedure proposed by Levy and Lieb. 
Similar to equation, the procedure begins by restricting the total energy evaluation to the class of many-body wavefunctions \{$\ket{\Psi}$\} that yield the same electron density $n({\bf r})$. Then, one can define a unique lowest energy for that electron density $n$ by varying $\ket{\Psi}$,
\begin{equation}
\begin{aligned}
    E_{\rm LL} &= \min_{\ket{\Psi} \in M(n)} \left[ \braket{\Psi | \hat{T} | \Psi} + \braket{\Psi | \hat{V}_{\rm int}| \Psi} \right] + \int d^3 r\, v_{\rm ext}({\bf r})\, n({\bf r}) + E_{II} \\
    &=F_{\rm LL}[n] + \int d^3 r\, v_{\rm ext}({\bf r})\, n({\bf r}) + E_{II},
\end{aligned}
\end{equation}
where $M(n)$ is the set contains all those many-body wavefunctions, $\ket{\Psi}$, which yield the density $n$. $F_{\rm LL}$ is the Levy-Lieb functional, once its minimum energy has been determined, the second step is to vary the electron density $n({\bf r})$ in order to minimize the total energy functional $E_{\rm LL}$. This process continues until converges to the ground state electron density $n({\bf r})$, along with its wavefunction $\Psi^{(1)}$ and total energy $E^{(1)}$. It follows immediately 
\begin{equation}
    E^{(1)} = \bra{\Psi^{(1)}}  \hat{H} \ket{\Psi^{(1)}} < \bra{\Psi_{\rm trial}}  \hat{H} \ket{\Psi_{\rm trial}}.
\end{equation}
This instructive formulation clarifies the meaning of the functional and provides an operational definition, while also removing the restriction to non-degenerate ground states present in the original Hohenberg-Kohn proof.

Of particular importance corollary of the Hohenberg Kohn theorem is that, since the Hamiltonian is uniquely determined by the ground-state electron density. Consequelty, the wavefunction of any state including both the ground state and excited states [Mermin 1964], can, in principle, be obtained by solving the Schr\"odinger equation~\ref{Schrodinger equation}  with the Hamiltonian. In this context (As such), all properties of the system are unique functionals of the ground-state electron density $n({\bf r})$.


\subsection*{Data Description}

In our work, we randomly sampled dataset from MP\cite{jain2013commentary}. The dataset was partitioned into non-magnetic (NM, $\mu=0$) data (containing 5590 samples) and magnetic (M, $\mu \neq 0$) data (containing 2122 samples). From the NM data, we extracted samples containing phonon spectrum (PS) data (containing 638 samples). The three different types dataset include electronic density distributions along with both regression and classification target variables, with all missing values (NaN) systematically removed.



The 3D electronic density matrix data from the MP database were obtained via DFT calculations using VASP\cite{kresse1993ab,kresse1994ab,kresse1996efficiency}. As shown in Fig. 1(a), this 3D matrix data was transformed into 3D image format to better meet the requirements of 3D CNN models. For property regression and classification predictions using 3D image-format data from the MP database, relying solely on total electronic density as a single-feature input proved ineffective. In contrast, a four-feature data incorporating total electronic density ($\rho_{\text{total}} = \rho_{\uparrow} + \rho_{\downarrow}$), magnetization density ($\rho_{\text{mag}} = \rho_{\uparrow} - \rho_{\downarrow}$), spin up ($ \rho_{\uparrow} $), and spin down ($\rho_{\downarrow}$) data was transformed into 3D image format. Some material properties are selected for regression predictions, such as $\text{B}$, $\upsilon $, $\text{E}_\text{form}$, $\text{E}_\text{total}$, $\text{E}_\text{gap} $, $\text{E}_\text{fermi} $, $\text{W}_\text{band}$ and $\rho$. And Materials were classified by both electronic structure (gapped vs. gapless states) and phonon stability (with/without imaginary frequencies). NM dataset is used for regression predictions of $\text{B}$, $\upsilon$, $ \text{E}_\text{form} $, $\text{E}_\text{total} $, $\text{E}_\text{gap} $, $\text{E}_\text{fermi}$, $\text{W}_\text{band}$, and $\rho$. M dataset is used for predicting magnetic moment. PS dataset is used for classification predictions of phonon stability (no imaginary/imaginary). $\rho_{\text{mag}}$ feature was not considered due to its lack of meaningful variation, which fails to provide sufficient information for predictive purposes.

The division of datasets is based on the characteristics of the properties and the availability of data. The NM dataset includes materials with zero magnetic moment, thereby minimizing the influence of magnetic properties on the prediction of target properties. The M dataset comprises materials with nonzero magnetic moments and is specifically tailored for predicting magnetic properties that are directly related to the materials' magnetism, such as magnetic moment. The PS dataset encompasses materials with phonon spectrum data and is utilized to identify imaginary frequencies. This division enhances the specialization of predictions by focusing on the properties and data relevant to each type of prediction.

\subsection*{Similarity-based target grouping}

To explore the statistically correlated behaviors among regression targets and facilitate multi-task learning\cite{hu2019multi}, we performed similarity-based grouping of target properties. Specifically, we calculated the pairwise $r$ between all target properties using the following formula\cite{pearson1895vii}:

$$
r = \frac{\sum_{i=1}^{n}(X_i - \overline{X})(Y_i - \overline{Y})}{\sqrt{\sum_{i=1}^{n}(X_i - \overline{X})^2}\sqrt{\sum_{i=1}^{n}(Y_i - \overline{Y})^2}}
$$

where $X_i$ and $Y_i$ are the observed values of two target properties, and $\overline{X}$ and $\overline{Y}$ are their respective means. For each pair of target properties, we also computed the statistical significance ($p$-value) of the correlation. If the $p$-value for a pair of target properties is below the threshold, $r$ is retained as the similarity, otherwise, $r$ is set to zero to eliminate noise, resulting in an weighted similarity matrix $S \in [-1,1]^{8 \times 8}$. And this filtering ensures that only statistically meaningful relationships are considered in subsequent analysis. To further group the targets, we transformed the correlation matrix into a distance matrix $D \in [0,1]^{8 \times 8}$ using the formula $D = (1 - S)/2$, ensuring that the distance is non-negative and reflects both positive and negative correlations\cite{xie2012multi}. Hierarchical agglomerative clustering was then performed on this distance matrix using the complete linkage method, as implemented in the Scikit-Learn library. The number of clusters was set to three, and the resulting group assignments were used to partition the target properties into distinct clusters.








\subsection*{Training}

The data was divided into training, validation, and test sets in an 8:1:1 ratio. To further enhance the model's generalization capability, we implemented a data augmentation strategy comprising two core operations: We add Gaussian noise to the input images to improve the robustness of the model, specifically applying Gaussian noise with a standard deviation of 0.001 to the input data. Additionally, We apply random rotations to the input images, which helps the model learn more invariant features, by performing 90-degree rotations (1-3 times) with a  $50\%$  probability. These operations were adopted because we understood they could help maintain the physical meaning of the data while simultaneously improving model robustness\cite{shorten2019survey}.

In terms of model optimization, we have designed an innovative learning rate scheduling strategy. The base scheduler, WarmupCosineLR, adopts a two-stage design: during the warm-up phase (the first 10 epochs), after numerous experiments, the learning rate increases linearly from  $10^{-6}$ to $\times 10^{-3}$ to ensure a smooth start to training; in the main training phase, the learning rate decays according to a cosine annealing schedule, which allows for fine-tuning while maintaining the benefits of a larger initial learning rate. To address the issue of learning stagnation during model training, we have developed the AdaptiveLRScheduler, which automatically decreases the learning rate by 0.2 (with a minimum of no less than $10^{-6}$) when the validation loss does not improve for five consecutive epochs.

We first conducted single-task training on three independent datasets (NM, M, and PS) to establish performance baselines. To adapt the MSA-3DCNN framework for single-task training, model adjustments were primarily focused on ensuring single-target output. During the training process, we employed task-specific loss functions\cite{wang2022comprehensive}: Mean Squared Error (MSE)\cite{gunst1977biased} for regression tasks and BCEWithLogitsLoss for classification tasks, combined with the AdamW optimizer\cite{loshchilov2017fixing} for efficient optimization. 

Next, we conducted multi-task training based on the NM dataset, which included regression prediction and classification prediction of properties to enhance the model's predictive efficiency. This approach was adopted because, in the field of physics, there are complex interrelationships among various material properties. To fully leverage these underlying interrelationships and optimize model performance, we employed a grouped multi-task strategy to adjust the model for outputs corresponding to different property groups. After grouping, the target values of each group were standardized to ensure that the loss functions of different tasks were optimized on the same scale, preventing the loss value of one task from overshadowing or neglecting the learning process of other tasks due to its magnitude. During the training process, the loss functions used for regression and classification predictions were the same as those used in single-task training. Given that regression and classification tasks are heterogeneous with differing output scales, we weighted the losses of the two tasks when calculating the total loss. Specifically, the regression task was assigned a loss weight of 0.9, while the classification task was given a weight of 0.1. This weighting method was adjusted based on the proportion of tasks and data characteristics to achieve the best model training results.

\section*{Data availability}

The data that support the results of this study are available from the corresponding author upon reasonable request.

\section*{Code availability}

\bibliography{sn-bibliography}

\section*{Acknowledgements}
This work was supported by the Science and Technology Development Fund from Macau SAR (Grant No. 0062/2023/ITP2 and No. 0016/2025/RIA1) and the Macao Polytechnic University (Grant No. RP/FCA-03/2023). X.C. acknowledges financial support from the National Natural Science Foundation of Shandong Province (Grant No. ZR2024QA040). B.S. acknowledges financial support from Swedish Research Council (grant no. 2022-04309 and grant No. 2018-07082).


\section*{Competing interests}
The authors declare no competing interests.

\end{document}